\documentstyle[prl,aps,multicol,epsf]{revtex}
\renewcommand{\narrowtext}{\begin{multicols}{2}
\global\columnwidth20.5pc}
\renewcommand{\widetext}{\end{multicols}
\global\columnwidth42.5pc} \multicolsep = 8pt plus 4pt minus 3pt

\input{psfig.sty}

\begin{document}

\draft

\title{Canted ground state in artificial molecules at high magnetic fields.}

\author{L. Mart\'{\i}n-Moreno}

\address{Departamento de F\'{\i}sica de la Materia Condensada, 
ICMA(CSIC), Universidad de Zaragoza, Zaragoza 50009, Spain.} 

\author{L. Brey}

\address{Instituto de Ciencia de Materiales de Madrid (CSIC), 
Cantoblanco, 28049, Madrid, Spain.}

\author{C. Tejedor}

\address{Departamento de F\'{\i}sica Te\'orica de la Materia
Condensada, Universidad Aut\'onoma de Madrid,  28049 Madrid, Spain.}


\maketitle

\begin{abstract}

We analyze the transitions that a magnetic field provokes 
in the ground state of an artificial homonuclear diatomic molecule.
For that purpose, we have performed numerical diagonalizations for a 
double quantum dot around the regime of filling factor 2. 
We present phase diagrams in terms of tunneling and Zeeman couplings, and 
confinement strength. We identify a series of transitions 
from ferromagnetic to symmetric states through a set of canted states 
with antiferromagnetic couping between the two quantum dots. 

\end{abstract}


\narrowtext

It is well known\cite{books}, that the ground state (GS) of 
homonuclear diatomic molecules, as $C_2$, $O_2$ or $N_2$, suffers a series of 
transitions when the distance between the nuclea is varied. Particularly 
interesting are the singlet-triplet transitions, which determine the 
magnetic properties of gases formed with those molecules. 
Since the intramolecule interatomic distance is, essentially, constant
it is not easy to 
control the physical mechanisms responsible for such a behavior. Recently,
double quantum dots (DQD), which can be 
thought of as artificial homonuclear molecules\cite{Rontani,Partoens}, 
have opened up new possibilities, that should allow both a better 
understanding of the physics of these systems and the tailoring 
of their magnetic properties, as some parameters, which are 
fixed in natural molecules, can be
continuously varied in DQD. For instance, it is possible to change 
potential barrier heights, thus varying the tunneling rate between 
the two artificial atoms, without altering the interdot distance,   
which determines the electron-electron interaction. Also, the interdot distance
is a parameter that can be externally controlled by appropriate design of
the nanostructure defining the DQD.

The use of an external magnetic field $B$ to produce transitions in the GS
is a particularly interesting tool to study artificial molecules:
apart from the effect of Zeeman energy, the magnetic field introduces 
an additional length scale (the magnetic length) that, 
even for moderate fields,
can be of the order of the molecular dimensions in these artificial systems.
In this work we analyze the symmetric-ferromagnetic transition which occurs
in DQD's with N electrons as the magnetic field is increased.
Some previous works\cite{Palacios,Hu} have dealt with DQD 
in magnetic fields, but in a totally different regime: there, the
spin degree of freedom was not included, i.e., those works only considered 
magnetic fields high enough for the system to be always in the
ferromagnetic phase.   

We restrict ourselves to {\it homopolar} molecules, i.e., DQD with N even where 
the electrons are three dimensionally confined: 
in the $z$ direction they can reside in either one 
of two layers (top (T) or bottom(B)), separated by a distance $d$,  
while in the $x-y$ plane electrons are confined 
by the presence of a two-dimensional parabolic potential 
$V(\vec{r})= m^* \omega_0 r^2 / 2$ ($m^*$ is the electron mass, $r$ is
the in-plane distance) and to a range of magnetic fields
such that the DQD has a filling factor close to $\nu=2$.
This last restriction is motivated
by the symmetric-ferromagnetic transitions found in infinite
multicomponent Quantum Hall (QH) systems close to filling factor $\nu=2$\cite{DP}. 

We work in the symmetric gauge, with a
set of states restricted to the lowest orbital Landau level, i.e. 
single particle wavefunctions without nodes in the radial direction. 
In this basis set, the Hamiltonian of the DQD is
\begin{eqnarray}
&& H \! = 
\alpha M -{\Delta_z} S_z
-\frac{\Delta_{sas}}{2}\sum_{m \sigma}\left( c^\dagger_{m \sigma T}
c_{m\sigma B}+   c^\dagger_{m \sigma B}c_{m\sigma T} \right)
\nonumber \\
&& + \! \sum_{\{ m_j \} \sigma \sigma ' \Lambda \Lambda ' }
\frac{V^{\Lambda  \Lambda ' }_{m_1m_2m_3m_4}}{2}
c^\dagger_{m_1\sigma \Lambda  }c^\dagger_{m_2\sigma ' \Lambda '}
c_{m_3\sigma ' \Lambda '} c_{m_4\sigma \Lambda },
\label{hamilt}
\end{eqnarray}
where $\omega_c = eB/m^*$ and 
$\alpha=\hbar ([\omega_c^2+4\omega_0^2]^{1/2}-\omega _c)/2$. 
$m$ and $M$ are the third components of single particle and total
angular momentum respectively.
$\sigma$ and $S_z$ are the third components of single particle and 
total spins respectively and 
$\Lambda$ a layer index, equal to T or B. 
$\Delta _z=g\mu_BB$ is the Zeeman coupling with $g$ 
being the Land\'e $g$-factor and $\mu _B$ the Bohr magneton. 
$\Delta _{sas}$ is the single particle energy gap between
symmetric ({\it s}) and antisymmetric ({\it as}) combinations of T and B single
particle states. $V^{\Lambda \Lambda }$=$e^2/(\varepsilon r)$  and
$V^{ \Lambda \Lambda '}$=$e^2/(\varepsilon \sqrt {r^2+d^2})$ for
$\Lambda \! \neq \! \Lambda '$ are Coulomb interaction potentials,
$e$ being the electron
charge and $\varepsilon $ the dielectric constant.
In all our results, the energies are
given in units of $e^2/(\varepsilon l_B)$ with $l_B=[\hbar /(m^*[\omega _c^2+
4\omega _0^2]^{1/2})]^{1/2}$ being the magnetic length.
The eigenstates of the DQD are a function 
$d$, $\Delta _z$, $\Delta _{sas}$ and $\alpha$.
The spectrum is separated in subspaces labeled by the quantum numbers
$(M,S_z,P_I)$ where
$P_I=(-1)^{I/2}$ is the parity of the isospin
$I=N_s-N_{as}$ with $N_{s}$, $N_{as}$ being the number of electrons in
symmetric and antisymmetric states respectively.
It should be noted that, for $d\ne 0$, interdot and intradot 
Coulomb interactions are different. 
In this case, electron-electron interaction mixes states with 
different isospin, and $I$ is not a good quantum number.
However, the parity $P_I$ is always a good quantum number because 
it is related to a symmetry operation of the problem: 
the reflection $\cal{R}$ with
respect to the midplane between the two layers.
The application of $\cal{R}$ to any state (not necessarily an eigenstate)
with a well defined isospin would result in
the same state multiplied by $(-1)^{N_{as}}=(-1)^{(N-I)/2}$ which, apart
from the unessential constant $(-1)^{N/2}$, is precisely $P_I$ 
times the state. This shows that $\cal{R}$ applied to any state without a
well defined isospin would still produce the same state (except for a 
sign), provided the initial state had weight in different isospin subspaces
with the same isospin parity.
There is another symmetry operation in the problem: the inversion with
respect to the midpoint between the centers of the two quantum dots (QD). 
We do not pay special attention to it in the discussion of our results, 
because it does not add any information of relevance.

We have diagonalized numerically the Hamiltonian (\ref{hamilt}) 
for $N=6$ and $N=8$. As similar structure is found in both
cases, only results for $N=8$ are presented here. 

The analysis of the results will be helped by the
large experience accumulated in the knowledge of the GS of a 
multicomponent QH systems\cite{DP}.
When electrons are confined in a double layer (DL) in the regime of  
global filling factor $\nu =2$, they have as degrees of freedom
$\sigma $ ($\equiv \uparrow, \downarrow$) and layer index $\Lambda$.
Those degrees of freedom provoke a rich phase diagram in terms of tunneling and 
Zeeman couplings. Obviously, when Zeeman splitting $\Delta _z$ 
is much larger than tunneling splitting $\Delta _{sas}$, the system prefers 
a ferromagnetic GS $|F\rangle $ in which electrons occupy 
all the symmetric and antisymmetric states with spin $\uparrow$. In the 
opposite limit $\Delta _{sas}>>\Delta _z$, the GS $|S\rangle $ corresponds 
to electrons fully occupying the symmetric states with both $\uparrow$ and 
$\downarrow$ spins. The character of the transition between these two extreme 
situations is controlled by electron-electron interaction effects in the 
four-dimensional space of degrees of freedom. A very attractive 
proposal was made\cite{Zheng,DasSarma1,DasSarma2} for a GS, labeled 
as canted, which connects continuously between the 
$|F\rangle $ and $|S\rangle $ limiting cases. The canted state 
is a ferromagnet in the field direction while, for the direction perpendicular 
to the magnetic field, spins in different
layers have antiferromagnetic correlations. 
The properties of this state has been studied by a microscopic 
Hartree-Fock theory, a long wavelength field theory based on the 
quantum O(3) nonlinear sigma model and a bosonic spin theory
\cite{Zheng,DasSarma1,DasSarma2,Demler,MacDonald,Brey,Yang,Demler2}. 
Moreover, exact numerical diagonalizations for a small number of electrons in
a DL with spherical shape\cite{Schliemann} show that, in a translationally
invariant system, the canted phase 
survive quantum fluctuations; although the domain, in the
$(\Delta_z, \Delta_{SAS})$ parameter space, in which it is the GS    
is much narrower than what Hartree-Fock approximation predicts. 
From the experimental side, such a 
phase is consistent with the available information in DL by 
inelastic light scattering\cite{Pellegrini}, 
magnetoresistance\cite{Sawada} and capacitance spectroscopy\cite{Khrapai}. 

In our calculations in DQD, we also find that, for large $\Delta _z$ and moderate  
$\Delta _{sas}$, the DQD has a $|F\rangle $ GS while for large $\Delta _{sas}$
and moderate $\Delta _z$ the GS is $|S\rangle $. As in the infinite system, 
the DQD presents a transition between these two regimes through 
a set of more complicated states. 

In order to identify a GS obtained from numerical 
diagonalizations as a canted state, a difficulty arises from the fact 
that the eigenstates of the DQD have a well defined third component $S_z$ of 
the spin while the mean field wave function $|C^{MF}\rangle$ for 
the canted state is not an eigenstate of $S_z$. 
However, we can restore the broken symmetry of the 
mean field states by projecting on subspaces with well defined $S_z=N/2-n$:
\begin{eqnarray}
|C_n^{MF}\rangle= \int d \phi e^{i\phi n}|C^{MF}(\phi )\rangle
\label{projection}
\end{eqnarray} 
where $n$ is an integer number. In Eq. (\ref{projection}), 
$\phi$ is the angle defining a particular canted state $|C^{MF}(\phi )\rangle= 
\left( \Phi _{1-}(\phi),\Phi _{2-}(\phi) \right)$ in which, for filling factor 
2, the electrons are occupying two type of states   
\begin{eqnarray}
\Phi _{1-}(\phi)= \left( \begin{array}{c}
cos (\theta _1 /2) \\ -e^{i \phi}sin(\theta _1 /2) \\ 0 \\ 0 \end{array} \right)
\, ; \, \Phi _{2-}(\phi)=\left( \begin{array}{c} 0 \\ 0 \\ 
e^{i \phi}cos(\theta _2 /2) \\ -sin(\theta _2 /2) \end{array} \right) .
\label{canted}
\end{eqnarray}
The notation is the one used by Das Sarma {\em et al.} \cite{DasSarma2}
with the states (\ref{canted}) given in the basis of $(s,\uparrow)$, 
$(as,\downarrow)$, $(s,\downarrow )$, $(as,\uparrow)$. The angles 
$\theta _i$ depend on the Hamiltonian parameters being $\theta _1=\theta _2=0$ 
for the ferromagnetic state $|F\rangle$ and $\theta _1=0$ and $\theta _2=\pi$
for the symmetric state $|S\rangle $.
After some algebra, one gets the projections (\ref{projection}) of the canted 
state written as  
\begin{eqnarray}
|C_n^{MF}\rangle& = & {\cal{C}}_n
\left[ cos (\theta _1/2) cos (\theta _2/2) 
\sum _{m} c^\dagger_{m, as, \downarrow }c_{m,s,\uparrow }
\right. \nonumber \\ & & \left. 
+ sin (\theta _1/2) sin (\theta _2/2)
\sum _{m} c^\dagger_{m, s, \downarrow }c_{m, as, \uparrow }
\right]^n|F\rangle 
\label{operator}
\end{eqnarray} 
${\cal{C}}_n$ being a normalization constant. 
The coefficients in Eq.(\ref{operator}) are 
$m$-independent in the infinite system because all single particle states 
are degenerate in the lowest Landau level. However, in a parabolic QD, 
translational invariance is broken, the single particle 
energy depends on $m$ and the corresponding canted eigenstates of the DQD 
should have a structure similar to Eq.(\ref{operator}) although the 
coefficients $\theta _1^m$ and $\theta _2^m$ could have a dependence on $m$.

\begin{figure}
\psfig{figure=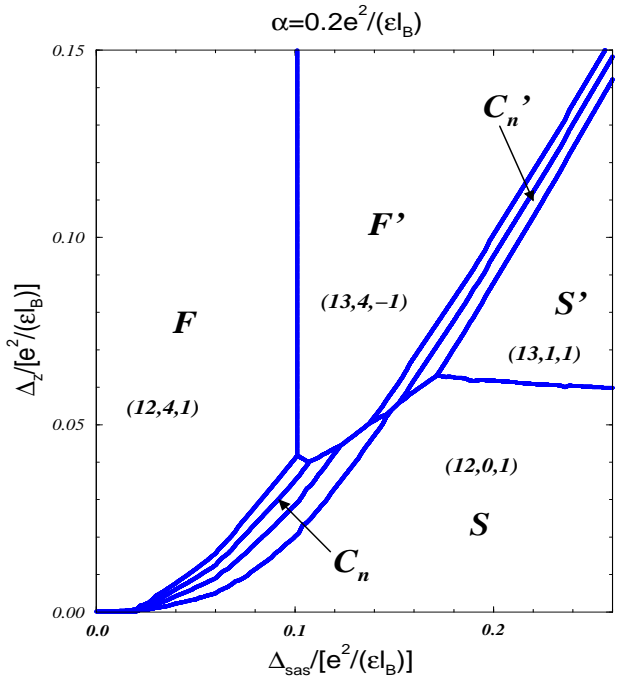,height=9cm,width=8cm}
\caption{Phase diagram showing the subspace $(M,S_z,P_I)$ containing the GS 
for $d=l_B$, $\alpha =0.2e^2/(\varepsilon l_B)$ and $N=8$. 
Energies $\Delta _{sas}$ and $\Delta _z$ are given in units of 
$e^2/(\varepsilon l_B)$.}
\label{fig1}
\end{figure}

Fig. \ref{fig1} contains our results in a phase diagram showing the quantum 
numbers $(M,S_z,P_I)$ of the GS for $d=l_B$ and $N=8$. In order to work in 
the regime $\nu =2$, we chose $\alpha =0.2e^2/(\varepsilon l_B)$.
In the region to the left corresponding to small $\Delta _{sas}$, the GS,
labeled as (12,4,1) is a ferromagnetic state given by a single 
Slater determinant $|F\rangle=\prod_{m=0}^3 c^\dagger 
_{m,s,\uparrow} c^\dagger_{m,as,\uparrow}$.
There is another large ferromagnetic region labeled as (13,4,-1) with a 
GS which is practically (more than $98 \%$) one Slater determinant 
$|F'\rangle=\prod_{m=0}^4 
c^\dagger_{m,s,\uparrow} \prod_{m=0}^2 c^\dagger_{m,as,\uparrow}$. This means 
that for an increasing tunneling, the ferromagnetic state $|F\rangle$ 
suffers an edge reconstruction by a charge instability which preserves the 
ferromagnetic character giving $|F'\rangle$.
In the lower region corresponding to small $\Delta _z$, the GS, 
labeled as (12,0,1), is a symmetric state given by practically (more than $98 \%$) 
one Slater determinant $|S\rangle=\prod_{m=0}^3 c^\dagger 
_{m,s,\uparrow} c^\dagger_{m,s,\downarrow}$. When the Zeeman coupling increases,
there is a new symmetric GS, labeled as (13,1,1), which is also practically
given by $|S'\rangle=\prod_{m=0}^4c^\dagger_{m,s,\uparrow}\prod_{m=0}^2c^\dagger_
{m,s,\downarrow}$. The increase of $\Delta _z$ provokes an edge reconstruction 
which involves a charge-spin excitation giving a new GS which essentially 
preserves the symmetric character. 

The most interesting part of the phase diagram corresponds to the narrow 
regions separating ferro-like regions from symmetric-like regions. The 
regions (12,3,-1), (12,2,1) and (12,1,-1) contain canted states 
$|C_n\rangle$ (corresponding to $n=1,2,3$ respectively) 
while the regions (13,3,1) and (13,2,-1) contain canted states 
$|C'_n\rangle $ (corresponding to $n=1,2$ 
respectively). This identification is made through the wavefunctions which 
turn out to have the functional form given by 
Eq. (\ref{operator}) with coefficients 
$\theta _1^m$ and $\theta _2^m$ almost independent on 
$m$ at the center of each QD. Since these regions are well described by 
states (\ref{operator}), they must be understood as the projection to 
well defined quantum numbers of canted states with an antiferromagnetic 
tilting of the spins. 

\begin{figure}
\psfig{figure=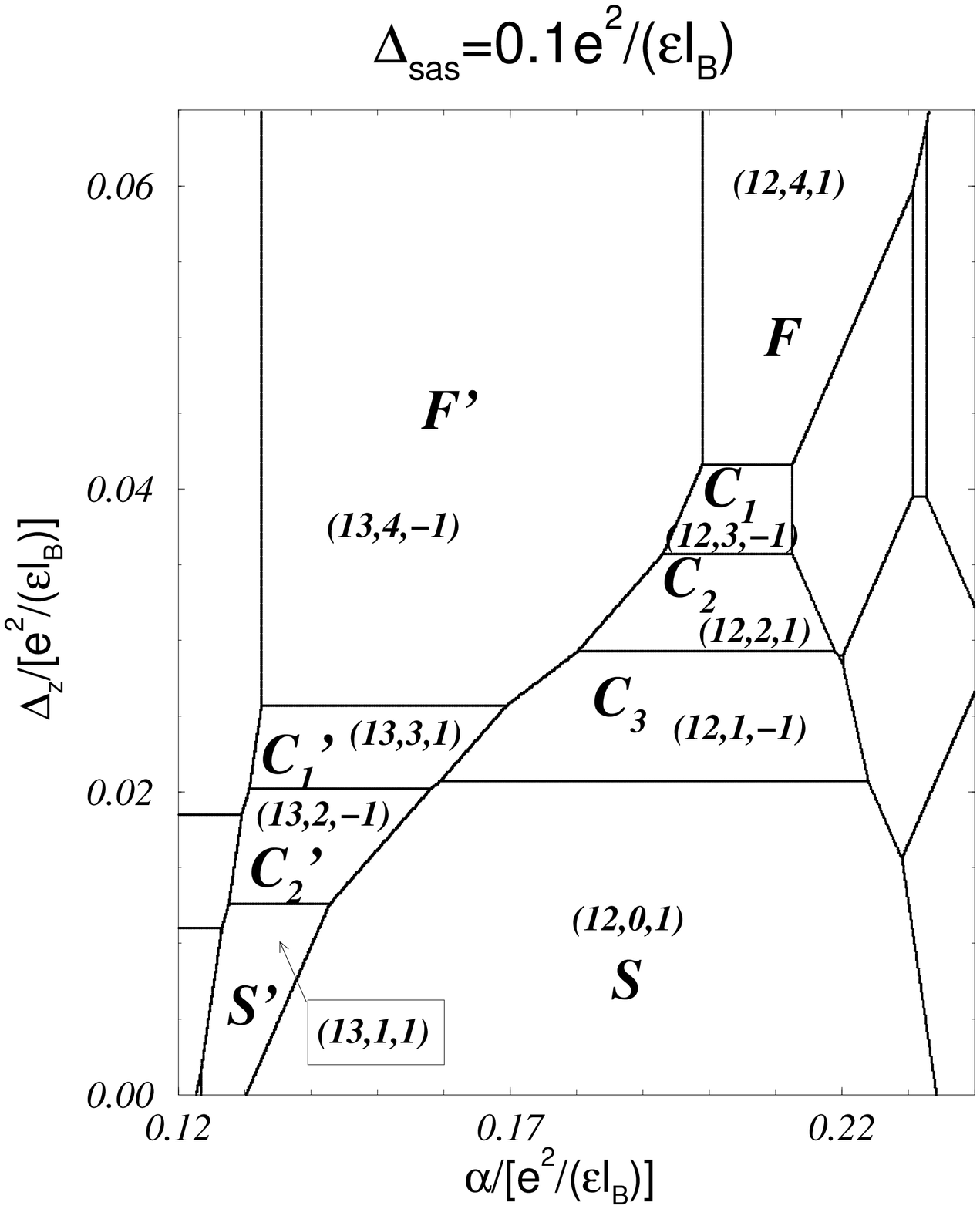,height=9cm,width=8cm}
\caption{Phase diagram showing the subspace $(M,S_z,P_I)$ containing the GS 
for $d=l_B$, $\Delta _{sas}=0.1e^2/(\varepsilon l_B)$ and $N=8$. 
Energies $\alpha$ and $\Delta _z$ 
are given in units of $e^2/(\varepsilon l_B)$.}
\label{fig2}
\end{figure}

\begin{figure}
\psfig{figure=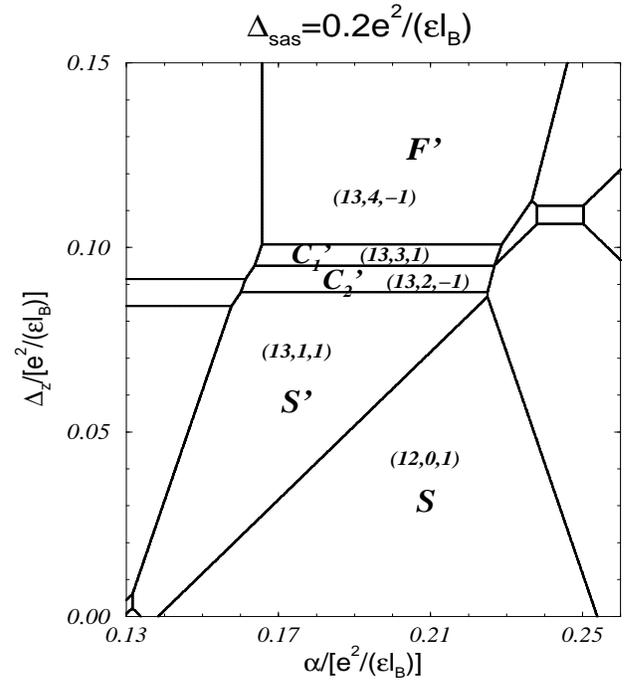,height=9cm,width=8cm}
\caption{Phase diagram showing the subspace $(M,S_z,P_I)$ containing the GS 
for $d=l_B$, $\Delta _{sas}=0.2e^2/(\varepsilon l_B)$ and $N=8$. 
Energies $\alpha$ and $\Delta _z$ 
are given in units of $e^2/(\varepsilon l_B)$.}
\label{fig3}
\end{figure}

Apart from the states shown in Fig. \ref{fig1}, some 
other GS, in particular with different edge reconstructions, could appear 
when the parameters are varied in the calculation. In fact this is the case:
as shown in figures \ref{fig2} and \ref{fig3}, showing phase diagrams 
for $d=l_B$, $N=8$, when $\Delta _z$ and $\alpha$ are varied 
for fixed $\Delta _{sas}=0.1e^2/(\varepsilon l_B)$ and 
$\Delta _{sas}=0.2e^2/(\varepsilon l_B)$ respectively.  
In the central part of Fig. \ref{fig2}, the sequence (12,4,1), 
(12,3,-1), (12,2,1), (12,1,-1) and (12,0,1) corresponds to the already 
discussed transition $|F\rangle \rightarrow |C_n\rangle 
\rightarrow |S\rangle$ with $n=1,2,3$.
Contiguously to its left, for smaller values of $\alpha$, 
the sequence (13,4,-1), (13,3,1), (13,2,-1)
and (13,1,1) corresponds to the  transition $|F'\rangle \rightarrow 
|C'_n\rangle \rightarrow |S'\rangle$ with $n=1,2$.
To the right of those regions, the increase in $\alpha $ implies 
a higher confinement and, consequently, a larger
concentration of the electrons in 
the center of the DQD which, in turns, implies a decrease of $M$. 
Edge reconstruction
occurs in the leftmost region, i.e., smallest $\alpha$, where a 
reduction of the confinement provokes 
the spreading of the electron density away from the center. 
In Fig. \ref{fig3}  the tunneling is so large that the sequence (12,4,1),
(12,3,-1), (12,2,1) and (12,1,-1) does not appear and only the symmetric 
state $|S\rangle \equiv (12,0,1)$ is observed. Apart from this, the whole 
structure is similar to that shown in Fig. \ref{fig2}.

In summary, we have analyzed the transitions that a magnetic field provokes 
in the GS of an artificial homonuclear diatomic molecule.
For that purpose, we have carried out numerical diagonalizations for a DQD 
for filling factors close to $\nu=2$. 
The resulting phase diagrams, are 
understood to the light of previous experience on the GS of an infinite
DL at the same regime. When the different parameters are varied, 
a series of transitions $|F\rangle \rightarrow |C_n\rangle
\rightarrow |S\rangle$ from ferromagnetic to symmetric states are identified,
through a set of projections (into subspaces with well defined quantum numbers) 
of canted states. Such canted states have ferromagnetic correlations in parallel 
(to magnetic field) direction and antiferromagnetic correlations 
in the perpendicular components of spins in different QD.
The electron-electron correlation that leads to the canted states in
infinite double layers for $\nu=2$ survives, in this regime, to the presence 
of edges and is present even when there is  edge reconstruction.

Work supported in part by the MEC of Spain under contract PB96-0085, 
the Fundaci\'on Ramon Areces and the CAM under contract No. 07N/0026/1998.


\widetext


\begin{references}
\bibitem{books} See for instance G. C. Lie and E. Clementi, J. Chem. Phys. 
{\bf 60}, 1288 (1974); R. McWeeny {\it Coulson's valence}, 
(Oxford University Press, Oxford, 1979); N. H. March and J. F. Mucci 
{\it Chemical physics of free molecules}, (Plenum Press, New York, 1993).
\bibitem{Rontani} M. Rontani {\it et al.}, Sol. State Commun., {\bf 112}, 
151 (1999).
\bibitem{Partoens}B. Partoens and F. M. Peeters, Phys. Rev. Lett.,
{\bf 84}, 4433 (2000).
\bibitem{Palacios}J. J. Palacios and P. Hawrylak, Phys. Rev. B, {\bf 51}, 
1769 (1995).
\bibitem{Hu} J. Hu {\it et al.}, 
Phys. Rev. B, {\bf 54}, 8616 (1996). 
\bibitem{DP}{\it Perspectives in Quantum Hall Effects} ed. S. DasSarma and 
A. Pinczuk, (Wiley, New York, 1997).
\bibitem{Zheng}L. Zheng {\it et al.}, Phys. Rev. Lett.,
{\bf 78}, 2453 (1997).
\bibitem{DasSarma1}S. Das Sarma {\it et al.}, Phys. Rev. Lett., 
{\bf 79}, 917 (1997). 
\bibitem{DasSarma2}S. Das Sarma {\it et al.}, Phys. Rev. B, {\bf 58}, 4672 (1998).
\bibitem{Demler}E. Demler and S. Das Sarma, Phys. Rev. Lett., {\bf 82}, 
3895 (1999).
\bibitem{MacDonald}A. H. MacDonald {\it et al.}, 
Phys. Rev. B, {\bf 60}, 8817 (1999).
\bibitem{Brey}L. Brey {\it et al.}, 
Phys. Rev. Lett., {\bf 83}, 168 (1999).
\bibitem{Yang}K. Yang, Phys. Rev. B, {\bf 60}, 15578 (1999).
\bibitem{Demler2} E. Demler {\it et al.},
Phys. Rev. B, {\bf 61}, R10567 (2000).
\bibitem{Schliemann}J. Schliemann and A. H. MacDonald, Phys. Rev. Lett.,
{\bf 84}, 4437 (2000). 
\bibitem{Pellegrini}V. Pellegrini {\it et al} Phys. Rev. Lett. {\bf 78}, 310 (1997);
Science, {\bf 281}, 799 (1998). 
\bibitem{Sawada}A. Sawada {\it et al} Phys. Rev. Lett. {\bf 80}, 4534 (1998). 
\bibitem{Khrapai}V. S. Khrapai {\it et al}, Phys. Rev. Lett. {\bf 84}, 725 (2000). 









\end{references}
\end{document}